\documentclass[12pt,preprint]{aastex}

\slugcomment{LA-UR-04-0006}
\shorttitle{Interiors of Jupiter and Saturn}
\shortauthors{Saumon and Guillot}
\def\wig#1{\mathrel{\hbox{\hbox to 0pt{%
        \lower.5ex\hbox{$\sim$}\hss}\raise.4ex\hbox{$#1$}}}}
\def\sss{\scriptscriptstyle}

\begin{document}

\title{Shock Compression of Deuterium and \\
the Interiors of Jupiter and Saturn}

\author{D. Saumon\altaffilmark{1}}
\affil{Los Alamos National Laboratory,  MS F699, Los Alamos, NM 87545, USA} 
\altaffiltext{1}{Dept. of Physics \& Astronomy, Vanderbilt University, Nashville, TN, 37235-1807}
\email{dsaumon@lanl.gov}

\and
\author{T. Guillot}
\affil{Observatoire de la C\^ote d'Azur, BP 4229, 06304 Nice CEDEX 04, FRANCE}
\email{guillot@obs-nice.fr}

\begin{abstract}
Recently, deuterium has been the focus of a high level of 
experimental and theoretical activity that was sparked 
by a disagreement on the experimental value of the maximum 
compression along the Hugoniot.  The behavior of deuterium 
at Mbar pressures is not well understood.

It is of great interest to understand 
how the current uncertainty on the hydrogen/deuterium EOS 
affects the inferred structures of Jupiter and Saturn.  In 
particular, the mass of a core of heavy elements (other than 
H and He) and the total mass of those heavy 
elements in these two planets are quite sensitive to the 
EOS of hydrogen and constitute important clues to their 
formation process.  We present a study of the range of 
structures allowed for Jupiter and Saturn by the current 
uncertainty in the hydrogen EOS and astrophysical observations 
of the two planets.  An improved experimental
understanding of hydrogen at Mbar pressures and better
determinations of the gravitational moments of both planets
are necessary to put tight bounds on their internal structure.
\end{abstract}


%%%%%%%%%%%%%%%%%%%%%%%%%%%%%%%%%%%%%%%%%%%%
%% MAINMATTER
%%%%%%%%%%%%%%%%%%%%%%%%%%%%%%%%%%%%%%%%%%%%

\section{Introduction}

Dense hydrogen has been the subject of numerous theoretical and
experimental studies over the last half century, in part because it
has the simplest electronic structure of all the elements and because
of its prevalence in the universe.  This simplicity is only
superficial however, as multiple experiments using static and dynamic
compression to Mbar pressures have revealed increasingly complex
properties and a rich phase diagram.  For example, there are three known
phases for the molecular solid \citep{mao94}, the molecular fluid
becomes conducting at $P\wig> 1.4\,$Mbar and $T \sim 4000\,$K (Weir, Mitchell 
\& Nellis 1996), the molecular nature of hydrogen persists to pressures
above 3$\,$Mbar (Loubeyre, Occelli \& LeToullec 2002), and superfluid and superconducting
phases have been suggested \citep{ashcroft03}.  Theoretical and
experimental research on hydrogen is usually motivated by interest in
warm dense matter problems in condensed matter physics, and 
applications to inertial confinement fusion research and to the
interiors of giant planets.  Indeed, Jupiter and Saturn are the
largest reservoirs of fluid metallic hydrogen in the solar system and
our understanding of these two planets depends rather critically on
the properties of hydrogen, especially its equation of state (EOS).

Jupiter and Saturn are composed of about 50\% to 70\%
hydrogen by mass.  At pressures of $P \sim 1 - 3\,$Mbar (corresponding
to $\sim 80$\% and $\sim 60$\% of the planet's radius for Jupiter and
Saturn, respectively), hydrogen goes from an insulating molecular
fluid to an atomic metallic fluid.  This transition has been the
subject of much theoretical and experimental work but remains poorly
understood.  Recent shock compression experiments of deuterium to Mbar
pressures disagree significantly \citep{collins98,knudson04,belov02}.

The internal structure of jovian planets 
is inferred indirectly from their global properties.  Their rapid rotation results
in a noticeable deformation and a non-spherical gravitational field
that can be expressed as an expansion in Legendre polynomials:
\begin{equation}
V(r,\theta)=-{GM \over r}\Bigg[1 - \sum^\infty_{n=1} \bigg({R_{\rm eq}
\over r}\bigg)^n J_n P_n(\cos \theta)\Bigg],
\end{equation}
where $G$ is the gravitational constant, $M$ the mass of the planet,
$R_{\rm eq}$ its equatorial radius, and the coefficients $J_n$ are
given by
\begin{equation}
J_n=-{1 \over MR_{\rm eq}^n}\int_V {r^\prime}^n P_n(\cos \theta)
\rho(r^\prime,\theta)\,d^3r^\prime
\end{equation}
and are known as the gravitational moments.  The integral is performed
over the volume of the planet.  To a very high degree of accuracy, the
planet is in hydrostatic equilibrium and the symmetry between the
northern and southern hemispheres implies that $J_n=0$ for odd $n$.
The first three non-vanishing moments, $J_2$, $J_4$, and $J_6$ have
been measured during spacecraft flybys of both planets.  Combined with
the total mass, the radius, and the rotation period of the planet,
these provide integral constraints on the density profile of the
planet $\rho(r,\theta)$.  These constraints cannot be inverted to
obtain the density profile, however.  Instead, a simple model is
assumed for the structure and composition of the rotating planet which
is then subjected to the condition of hydrostatic equilibrium,
including the rotational potential perturbation.  Model parameters are
adjusted to fit the observed constraints.  The EOS provides the
$P(\rho)$ relation needed to close the system of equations.  The
structure thus inferred is sensitive to the choice of hydrogen EOS
used in the models.

The total amount of heavy elements and their
distribution inside these two planets bear directly on their formation
process by accretion of both gaseous and solid material
from the protoplanetary nebula.  It is generally thought that Jupiter
and Saturn formed according to the core accretion model.  In the
regions where they formed, refractory compounds (most significantly,
water) had condensed and formed small solid bodies.  Collisional
processes between these planetesimals led to the build up of cores of
10 -- 20$\,M_\oplus\,$ \citep{pollack96} 
of heavy elements that subsequently accreted the
surrounding gas on a dynamical time scale to
build the large planets that we see today. \footnote{The total masses
are 317.83$\,M_\oplus$ and 95.147$\,M_\oplus$ for Jupiter and Saturn,
respectively.}  Formation by a disk instability in the gaseous
protoplanetary disk, which does not require a seed core of heavy
elements, has also been suggested as a formation mechanism \citep{boss00}.
Whether formation occurred through core accretion or
disk instability (or perhaps another mechanism) depends very much on
the internal structure and distribution of heavy elements that we see
today in these planets.

Since the last determination of the gravitational moments of Jupiter
and Saturn two decades ago \citep{campbell85, campbell89}, modeling of
the interiors of Jupiter and Saturn has been driven primarily by the
availability of improved equations of state for hydrogen.  Recently,
measurements of the reshock temperature (Holmes, Ross \& Nellis 1995), the
development of laser-induced shock compression \citep{dasilva97} and
pulsed-power shock compression \citep{knudson01} have provided a
wealth of new data and fueled intense theoretical debate on the
physics of hydrogen at Mbar pressures and temperatures of $\sim
10^4\,$K.  For the first time, there are experimental data that allow us
to constrain the EOS of hydrogen inside jovian planets at pressures of
$\wig> 1\,$Mbar, where the EOS remains most uncertain.  Experimental
results along the principal Hugoniot\footnote{A Hugoniot is a curve in
$(P,V,T)$ space that represents all the possible shocked states that can be
achieved from a
given initial state $(P_0,V_0,T_0)$.} of deuterium do not agree
\citep{collins98, knudson01} in this pressure range.  Laser
compression data give a maximum compression of $\sim 6\,$
\citep{dasilva97, collins98} while the pulsed-power compression
experiments (\cite{knudson01}) and convergent spherical shock wave experiments
\citep{belov02, boriskov03} find a value of $\sim 4\,$.  Further study
will reveal which experimental result is correct.  However, it is
generally agreed that the actual Hugoniot does not lie outside the
range allowed by the current experiments.  It is now possible to
assign realistic error bars on the principal Hugoniot of deuterium,
and hence on the EOS at pressures of a few Mbar.  Until now,
estimating the uncertainties on the EOS of hydrogen at these pressures
had been little more than guess work.

In view of these new developments, a new study of the interiors of
Jupiter and Saturn is warranted.  In this paper, we examine how the
uncertainty on the EOS of hydrogen maps unto uncertainties on the
interior structure of Jupiter and Saturn.  The resulting range of
models provide firm bounds on the amount and distribution of heavy
elements in these two planets.  We first generated a number of EOS,
supplemented with existing tabular EOS to represent subsets of
$(P,V,T)$ data along the principal and reshock Hugoniots of deuterium.
Together, these EOS give Hugoniots that reproduce the extremes found
in the data as well as intermediate behavior.  The EOS are described
in section 2.  The computation of
interior models of Jupiter and Saturn, described in section
3, includes many improvements
over previous work.  We present the new models and the effect of the
hydrogen EOS on the cooling time scale of Jupiter in sections
4 and 5, respectively. The astrophysical implications are discussed in section 6.

\section{Equations of State}

\subsection{Interior structures and relevant experimental data}

The interiors of Jupiter and Saturn are believed to be mostly
convective. The very high efficiency of convection inside jovian
planets leads to a very small superadiabaticity so that the planets'
interiors should be fully adiabatic for all practical purposes. This
assumption may break down in several regions: (i) near the planets'
cores, if conduction by degenerate electrons becomes efficient enough,
or if a gradient of molecular weight is present; (ii) in a region at
$\sim 1\,$Mbar where helium would become less soluble in metallic
hydrogen; (iii) at the molecular-metallic transition of hydrogen, if
it is discontinuous; (iv) at $\sim 1\,$kbar levels if alkali metals are not
present (i.e. if their abundances is significantly less than solar); (v)
in the ``meteorological layer'' ($P\sim 0.3-40$\,bar), where the
condensation of water affects the temperature gradients, chemical
composition and horizonal homogeneity. None of these possibilities can be
dismissed at present but it can be advocated that none of them may cause
a significant departure of the interior profile from 
adiabaticity \citep{gshs03}. In the absence of further information, 
adiabaticity of the interior profile is a perfectly valid working hypothesis.

The structure of the fluid envelopes of the giant planets is thus
fixed by their specific entropy determined from observations of the
surface.  On the other hand, shock-compression experiments follow
Hugoniots that are typically much hotter than the Jupiter and Saturn
adiabats for the pressures of interest.  For example, at 1$\,$Mbar the
temperature inside Jupiter is $\sim 6000\,$K, while the principal
Hugoniot reaches $\sim 20000\,$K.  The gas-gun reshock experiments
overlap Jupiter's adiabat up to 0.8$\,$Mbar, however
(Figs.~\ref{fig:adiab_Prho} and \ref{fig:adiab_PT}).

Because of the limited overlap between Hugoniot data and the jovian planet adiabats,
it is necessary to rely on model EOS anchored to the Hugoniot data to compute 
adiabats.  This implies that two EOS models that predict
nearly identical Hugoniots may produce different adiabats.  For the
purposes of this study, we computed interior models of Jupiter and
Saturn with 7 different hydrogen EOS.  These have been chosen as
representative EOS that reproduce selected subsets of data and
realistically bracket the actual EOS of hydrogen.

For simplicity, we consider only the following experiments:
\begin{itemize}
  \item
    The $(P,V)$ principal Hugoniot measured with the Z-machine at
       Sandia National Laboratory \citep{knudson01, knudson03,
       knudson04} and two Hugoniot points achieved by convergent
       spherical shock wave compression \citep{belov02, boriskov03}. These data
       indicate a maximum compression of $\sim 4$ for deuterium along
       the Hugoniot.
  \item
    The $(P,V)$ principal Hugoniot measured at the NOVA facility at
       Lawrence Livermore National Laboratory \citep{dasilva97,collins98}.
       These data indicate a maximum
       compression of $\sim 6$ for deuterium along the Hugoniot.
  \item
    The $(P,T)$ data along the principal Hugoniot measured at the NOVA
       facility \citep{collins01}.
  \item
    The single and double shock gas gun $T$ measurements \citep{holmes95}.
    The double shock temperatures may be
       systematically underestimated due to unquantified thermal
       conduction into the window upon shock reflection.  We consider
       this data set as a lower limit on the reshock temperatures.
  \item
    The single and double shock gas gun $(P,V)$ measurements \citep{vanthiel74, nellis83}.
\end{itemize}

Together, these experiments characterize the principal and reshock
Hugoniots of deuterium and the full range of current uncertainties quite well.
Shock reverberation \citep{knudson03} and laser reshock experiments to
several Mbar \citep{mostovych00, boehly04} have been shown to be
generally consistent with (or to fall in between) EOS that we consider
below, or similar EOS models.

\subsection{Equations of State for Hydrogen}

We have developed 4 different EOS based on the simple linear mixing
model (hereafter, LM) devised by Ross \citep{holmes95,ross98,ross99} to reproduce the
unexpectedly low gas-gun reshock temperature measurements \citep{holmes95}.
The linear mixing model is based on a linear
interpolation in composition between a molecular fluid and a metallic
fluid:
\begin{equation}
F=(1-x)F_{\rm mol} + 2x(F_{\rm met}+F_{\rm fit}) - TS_{\rm mix}(x), 
\end{equation}
where $F$ is the total Helmholtz free energy per 2 atoms, $F_{\rm
mol}$ and $F_{\rm met}$ are the free energies of the molecular and
metallic fluids, respectively, $x$ is the fraction of dissociated
molecules, and $S_{\rm mix}(x)$ is the ideal entropy of mixing.  The
molecular free energy is obtained from fluid perturbation theory with
a soft sphere reference potential and an effective H$_2$-H$_2$ pair
potential fitted to the gas gun $(P,V)$ data (Ross, Ree \& Young 1983).  All four EOS
presented here thus produce the same molecular Hugoniot.  The One Component Plasma
(OCP) forms the basis of the metallic fluid free energy, with a
zero-temperature electron EOS including exchange and 
correlation \citep{holmes95}.  An entropy term ($F_{\rm fit}=\delta_e kT$, where
$k$ is the Boltzmann constant) is introduced in the metallic free
energy and adjusted to fit the reshock temperature data
($\delta_e=-2.7$).  This model was found to agree well with the
subsequent NOVA Hugoniot \citep{collins98,dasilva97} with the same
value of $\delta_e$.  The LM model is particularly useful for the
present study as it can be easily modified to fit various data sets.

We found that the original LM model \citep{ross98,ross99} predicts an
anomalous adiabat for Jupiter.  This can be seen in a sequence of
adiabats with decreasing specific entropy (Fig. \ref{fig:J_ad_Ross}).
The turnover of the adiabat around 1$\,$Mbar has been reported before
(Nellis, Ross \& Holmes 1995) and is a direct consequence of fitting the reshock
temperature measurements.  At higher pressures, a cusp in the adiabat
develops at low entropies.  While this behavior is not strictly
prohibited thermodynamically, it is very suspicious.  It arises from
the constant $\delta_e k$ shift in entropy of the metallic adiabats in
the model.  This causes a mismatch between the molecular and metallic
adiabats that is significant when the entropy becomes small enough, as
in Jupiter and Saturn.  This means that while the original form of the
LM model reproduces the NOVA Hugoniot quite well, it is inadequate for
EOS calculations at temperatures well below that of the principal
Hugoniot.

We computed four LM EOS that produce better-behaved adiabats:

\begin{itemize}
\item
  {\bf LM-A:} This EOS is identical to the original LM model
\citep{ross98,ross99} except that the anomalous behavior of the adiabat has
been corrected by replacing the fitting term $\delta_e kT$ by a
temperature- and density-dependent contribution that is adjusted to
fit Hugoniot data.  The functional form is chosen for flexibility
and to mimic the qualitative density dependence of electron screening: 
\begin{equation}
{F_{\rm fit} \over kT}={-2.5 \over {1 + (0.8/V)^2}} \bigg({5000 \over
T}\bigg)^{-0.1},
\end{equation}
where $V$ is the specific volume in cm$^3$/mol, $T$ the temperature in
Kelvin and $F_{\rm fit}$ is the free energy per 2 atoms.  The
resulting EOS reproduces the NOVA (P,V,T) data and the reshock
temperatures very well.
\item
  {\bf LM-B:} This EOS is identical to LM-A except that a steeper
dependence on the specific volume $V$ is used in $F_{\rm fit}$:
\begin{equation}
{F_{\rm fit} \over kT}={-2.5 \over {1 + (1.5/V)^4}} \bigg({5000 \over
T}\bigg)^{-0.1}.
\end{equation}
The other parameters in Eq. (4) have been adjusted to fit the same data as with
LM-A and their optimal values remain unchanged.  In particular, the temperature 
dependance is tightly constrained by the NOVA $(P,T)$ data \citep{collins01}.
Higher powers of $V$ in Eq. (5) cause a sharp minimum in the dissociation fraction $x$ 
around 1.3$\,$cm$^3$/mol which is anomalous.
The Hugoniots obtained for this form of $F_{\rm fit}$
are essentially identical to those of LM-A but the
adiabat is stiffer and slightly hotter for $P\wig>2\,$Mbar.
\item
  {\bf LM-SOCP:} In this model, the OCP model used by Ross for the
metallic fluid is replaced by a more realistic screened OCP model
\citep{chabrier90, pot_chab00} because electron screening is important
in the regime found in the metallic envelopes of jovian planets.  The
electronic contributions to the free energy include finite temperature
effects.  In addition, we keep the entropy term of Ross and adjust it
to approach the Sandia Hugoniot with $\delta_e = 0.5$. This (positive)
value of $\delta_e$ is a compromise between fitting the low compressibility of the
Sandia Hugoniot and the NOVA temperature measurements.  The Sandia
Hugoniot can be fit with $\delta_e=2.0$ but the resulting principal
Hugoniot temperatures are $>2\sigma$ higher than the NOVA $(P,T)$
data.  This model overestimates the reshock temperatures by $\sim
30$\% and the maximum compression ratio of the Sandia and the spherical shock
wave experiments by $\sim$15\%.\footnote{The convergent spherical shock wave
experiment uses solid deuterium for its initial, unshocked state, with a density of 
$\rho_0=0.20\,$g/cm$^3$, compared to the liquid density of 0.17\,g/cm$^3$ 
used in the other experiments.
This results in a slightly denser Hugoniot (Fig.~\ref{fig:hugo_panels_A}) but 
the compression ratio ($\rho/\rho_0$) is very close to that of the Sandia 
experiment.}
It is also a good representation of DFT-GGA
simulations of deuterium \citep{lenosky00} for $T \wig> 10^4\,$K and
$\rho \wig> 0.6\,$g/cm$^3$, {\it i.e.} outside of the molecular regime
where these simulations miss the gas gun $(P,V)$ data.
\item
  {\bf LM-H4:} In a variation on the LM EOS model, the need for a
   fitting term in the original LM model is removed by introducing
   ${\rm D}_4$ chains as a new species \citep{rossyang01}.  This model
   considers a mixture of linear D$_4$ chains, D$_2$ molecules and a
   dissociated D$^+$ + $e^-$ fluid metal.  Instead of the pair-wise 
   linear mixing scheme used by \cite{rossyang01}, we apply a 3-component
   LM mixing scheme:
\begin{equation}
F=(1-x-y)F_{{\rm D}_4} + 2xF_{\rm mol} + 4yF_{\rm met} -TS_{\rm
mix}(x,y)
\end{equation}
where $F$ is the free energy per 4 atoms, $x$ and $y$ are the
fractions of dissociated and ionized D$_4$ chains, respectively, $S_{\rm mix}(x,y)$
is the ideal mixing entropy of the 3-component system, and $F_{\rm
met}$ is the screened OCP free energy, as in LM-SOCP.  This 
has a much larger effect than the choice of linear mixing scheme. The resulting
EOS reproduces the NOVA $(P,V,T)$ Hugoniot well but overestimates the
reshock temperatures by $\sim 20$\%.
\end{itemize}

In all cases, the internal partition function of the molecules is
calculated by an explicit sum over the known internal energy levels of
H$_2$ or D$_2$ as appropriate.  Similarly, proper isotope scaling was
applied for the approximate vibrational and rotational properties of
the chains \citep{rossyang01}.  These four EOS models were applied to
deuterium to compare with shock compression experiments and to hydrogen to
compute interior models of planets.
These EOS are not intended to constitute practical EOS
models for purposes other than this sensitivity study.  They are
rather simple representations of subsets of data that include enough
physics to allow a reasonable calculation of adiabats.

Even though it predates all the shock compression experiments that we
consider here, the {\bf SESAME} deuterium EOS 5263
\citep{kerley72, sesame} provides a fair representation of the Sandia
Hugoniot and we adopt it here.  For hydrogen, we adopt the SESAME EOS
5251 \citep{sesame}, which is the deuterium EOS scaled in density.
The SESAME 5251 table provides only $P(\rho,T)$ and $U(\rho,T)$.  We
computed the entropy by integrating the internal energy downward along
isochores from a high-$T$ isotherm obtained from the Saumon-Chabrier
EOS (Saumon, Chabrier \& Van Horn 1995).  The calculated entropy 
does not recover the limit of the ideal H$_2$ gas, however.  
This is a consequence of the non-scalability of the 
molecular internal energy and entropy.

We found that in the molecular region ($P<0.2\,$Mbar), the SESAME deuterium
Hugoniot is somewhat stiffer than indicated by the better measurements
made after it was developed \citep{nellis83}. This is partly to blame
for our inability to compute satisfactory models of Jupiter and Saturn
with this EOS (see section 4).
We have therefore patched the SESAME EOS in the molecular regime with
an EOS that reproduces the low-$P$ Hugoniot data such as any of the
above linear mixing  EOS.  The patch is introduced by smoothly switching from one
EOS to the other by applying the additive volume rule in the switching
region, as if we were mixing two different substances.  This preserves
thermodynamic consistency. The transition is located between 0.1\,kbar and
0.4$\,$Mbar.  We label this EOS {\bf SESAME-p}.  The resulting
Hugoniots are nearly identical to SESAME Hugoniots, except at
pressures below $\sim 50\,$kbar.  Because the SESAME entropy does
not recover the ideal H$_2$ gas entropy at low density but the molecular EOS
patch does, the SESAME-p adiabat is shifted to higher $T$ and lower
$\rho$ at pressures above 10$\,$kbar.

Finally, we computed models with the {\bf SCVH-I} EOS \citep{scvh}.
Like the SESAME EOS, it predates all experiments except the $(P,V)$
gas gun data which it reproduces by construction (like the four LM
EOS above).  It agrees better with the Sandia $(P,V)$ Hugoniot for
$P \wig< 0.7\,$Mbar and then shifts toward the NOVA Hugoniot at higher
pressures across a small region where it is interpolated between the
low-density and the high density EOS.  The reshock temperatures are
overestimated by $\sim 40$\%, but the agreement with the NOVA
temperatures is good.  This EOS has been used extensively in modeling
the interiors of Jupiter and Saturn
\citep{cshl,gcmg94,ggh97,guillot99,gz99} and it serves as a basis for
comparison.

Each EOS is compared with the various shock compression experiments in
Figs.~\ref{fig:hugo_panels_A} and \ref{fig:hugo_panels_B}.  This set
of 7 EOS shows a range of behaviors that is representative of the
possible range indicated by the experiments.  The corresponding
hydrogen adiabats for Jupiter are shown differentially with respect to
the SCVH-I adiabat in Figs.~\ref{fig:adiab_Prho} and
\ref{fig:adiab_PT}.  The Saturn adiabats are very similar.
Interestingly, the variation among the adiabats in $(P,\rho)$
(Fig.~\ref{fig:adiab_Prho}) is rather modest.  By considering both the
experimental error bars, the discrepancies between various experiments
and a range of EOS models, we can estimate the uncertainty in the
density along the Jupiter adiabat.  The error bar on the adiabat
density is vanishingly small at 1\,bar (where the gas is ideal), and remains below
0.5\% up to 1$\,$kbar.  It grows steadily from 2\% at 30$\,$kbar to
8\% at 6$\,$Mbar.  The uncertainty decreases to about 3\% at
100$\,$Mbar because the fluid is increasingly ionized and the EOS is
more easily modeled.  This small uncertainty on the $(P,\rho)$ relation
along the adiabat is large enough to
affect the interior structure of the models.  We find that the
temperature along the adiabat (Fig.~\ref{fig:adiab_PT}) is much more
sensitive to the choice of EOS model.  Variations are typically of
$\sim25\,$\% above 3$\,$Mbar and can be as large as 60\%.  This
affects the thermal energy content of the planet and the time it takes
to cool to its present state.  The LM-A and LM-B adiabats are
particularly cool at pressures above 1$\,$Mbar.

\subsection{Equations of State for Helium}

Helium globally accounts for about 10\% of the atoms in the jovian planets. 
Very little data and few EOS are
available for helium in the regime of interest for jovian planets.  
The is only one set of four Hugoniot measurements published
for helium \citep{nellis84}, reaching 0.6$\,$Mbar on the reshock.  Static compression
experiments reach similar pressures at room temperature \citep{loubeyre93}.
The EOS of helium in the 1--100$\,$Mbar range is therefore unconstrained by 
experiments.  In view of the recent developments on the EOS of hydrogen,
we should assume that the EOS
of helium at Mbar pressures and temperatures of several thousand degrees is
much more uncertain than that of hydrogen, a situation that is fortunately mitigated
by its relatively low abundance in jovian planets.  We investigate
the importance of this additional source of uncertainty on the structures of Jupiter
and Saturn using two different EOS.

\begin{itemize}
\item
  {\bf He-SCVH:}  This tabular EOS \citep{scvh} is based on fluid perturbation theory for the atomic 
   fluid, with an effective He-He pair potential adjusted to fit the Hugoniot data,
   and a screened OCP model for the fully ionized, dense plasma \citep{chabrier90}.
   Between these two regimes (roughly from 0.6 to 60$\,$Mbar), the EOS is smoothly 
   interpolated.  Thus, most of the mass of Jupiter and Saturn falls in the interpolated part 
   of the He-SCVH EOS.  This helium EOS has been used extensively in models of
   Jupiter and Saturn \citep{cshl,gcmg94,ggh97,guillot99,gz99}.
\item
  {\bf He-SESAME-p:}  The SESAME EOS 5761 for helium \citep{sesame} predates the
   Hugoniot data \citep{nellis84} but nevertheless reproduces them fairly well,
   although it is somewhat softer than the data indicates.  To generate the entropy
   for the calculation of adiabats, we integrated the internal energy table along 
   isochores as we did for the SESAME hydrogen EOS.  There are some small anomalies
   in the internal energy in the regime of Saha ionization that result in
   low-density entropies that deviate from the ideal gas limit for $T<T_{\rm ioniz}$
   by 4--5\%.  To remediate these limitations, we have patched the SESAME helium
   EOS for pressures below 10$\,$kbar with the He-SCVH EOS following the same procedure
   as for the SESAME-p hydrogen EOS.
\end{itemize}

Figure~\ref{fig:He_adiab} compares the two helium EOS along the $(P,T)$ path 
defined by the {\it hydrogen} adiabat for Jupiter. The latter is a good approximation 
to the actual adiabat for a jovian mixture of hydrogen, helium, and heavy elements.    
The SESAME EOS becomes softer for $P\wig>10\,$kbar and then significantly stiffer
above 1$\,$Mbar.  Not surprisingly, the difference between the two He EOS
is larger than the most extreme differences found in the H EOS (Fig.~\ref{fig:adiab_Prho}).

\section{The construction of interior models}

The method used to compute interior models of Jupiter and Saturn has
been presented elsewhere \citep{gcmg94, guillot99} and will only be
summarized here. Their interiors are assumed to consist of
three mostly homogeneous regions:
\begin{enumerate}
\item A central dense core of mass $M_{\rm core}$ to be determined and of
  unknown composition;
\item An inner helium-rich envelope, assumed to coincide with the
  metallic hydrogen region;
\item An outer helium-poor envelope, whose helium abundance is
  constrained by spectroscopic and {\it in situ} measurements
  of the atmosphere. 
\end{enumerate}
The presence of a central core is generally needed to fit the planets'
gravity fields. It is also qualitatively consistent with models in
which the giant planets are formed by first accretion of a central
seed of solids in the otherwise gaseous protosolar 
nebula \citep{pollack96}.  The division of the hydrogen-helium envelopes into
an inner and an outer part arises because the observed helium
atmospheric abundances in Jupiter (von Zahn, Hunten \& Lehmacher 1998) and Saturn
\citep{conrath00} are reduced compared to the protosolar value
\citep{bahcall95}.  It is speculated that hydrogen and
helium undergo a phase separation for $T\wig< 5000\,$K and pressure of
a few Mbar \citep{stevenson75}. The formation of helium-rich droplets
that sink deeper into the planet would explain the reduced atmospheric
helium abundance. It would also provide an additional source of energy
that is needed to understand Saturn's evolution, but could become a
problem for Jupiter as it would tend to prolong its cooling beyond the
age of the Solar System \citep{ss77b,fortney03,gshs03}.
In the absence of an abundant species
susceptible to separate from hydrogen (apart from helium), we assume
that all heavy elements are homogeneously mixed in the envelope. 

Given these simplifying assumptions, the only two parameters that we
seek are the mass of the central core $M_{\rm core}$ and the mass of
heavy elements mixed in the hydrogen-helium envelope, $M_{\sss Z}$.  The value of
these parameters are determined by an optimization technique aimed at
finding models that match the observational values of the planets'
equatorial radii $R_{\rm eq}$ and gravitational moments $J_2$ and
$J_4$ within their error bars \citep{gcmg94,guillot99}.  Due to large
experimental uncertainties, $J_6$ does not further constrain the models. 

The following uncertainties are taken into account:
\begin{enumerate}
\item The central core is assumed to be formed either from ``rocks''
  (e.g. silicates and iron) or from ``ices'' ( e.g. water, methane, ammonia);
\item The mass fraction of rocks in the envelope is varied between 0
  and 4\%;
\item The temperature at 1\,bar, which determines the specific entropy
   of the adiabatic structure, is between 165 and 170\,K for Jupiter and 
  between 135 and 145\,K for Saturn;
\item The ratio of helium to hydrogen abundance in the atmosphere is
  set to the in situ measurement for Jupiter \citep{vonzahn98} and the
  inferred spectroscopic measurement for Saturn \citep{conrath00}. The
  overall helium to hydrogen ratio in the entire planet is set equal
  to the protosolar value inferred from solar interior
  models (Bahcall et al. 1995). Both values are varied by $\pm
  1\sigma$;
\item The location of the transition from the inner helium-rich to the outer
  helium-poor region is varied between 1 and 7\,Mbar
  for Jupiter, and between 1 and 3\,Mbar for Saturn;
\item Jupiter and Saturn are assumed to either rotate on concentric cylinders
  following observed zonal winds at the surface, or to rotate as solid
  bodies \citep{hubbard82}.
\end{enumerate}

Contrary to previous work, we do not allow for compositional
discontinuities in the abundances of heavy elements in the envelope.
We also assume that the temperature profile is adiabatic, as any
radiative layer, if present at all, should be confined to a small
region \citep{gshs03}. The hydrostatic structure is computed with the
theory of rotating figures developed to the 4$^{\rm th}$ order in the
rotational perturbation.

We use the EOS for hydrogen and helium as described in Section 2. As
an improvement over the previous models, we explicitly take the EOS of
rocks and ices into account. We use the SESAME EOS 7154 of water \citep{sesame} 
as a proxy for all ices (H$_2$O being by far the
most abundant) and the SESAME EOS 7100 of dry sand \citep{sesame} for
the rock-forming heavy elements. We account for the
condensation of rocks/silicates in an approximate fashion
by setting their abundance to zero for
temperatures $T<2500\,$K. Water condenses around 300\,K, but this is not
included as it has a negligible effect on the interior structure and
gravitational moments.  In the absence of a suitable theory,
the EOS of mixtures of H, He, water, and dry sand 
is obtained by applying the additive volume rule (see \S 4.1,
however).

The 3-layer model is simplistic and the interiors of
Jupiter and Saturn are undoubtedly more complex.  Due to the scarcity
of data and its current level of precision, more elaborate models are
not justified at this point, however. We believe that the most
significant sources of uncertainty have been taken into account in
this study, which represents the most extensive sequence of interior
models computed for Jupiter and Saturn so far.

\section{New optimized interior models}

\subsection{Jupiter}

For each EOS considered, a range of core masses ($M_{\rm core}$) and
masses of heavy elements in the envelope ($M_{\sss Z}$) is obtained
after varying the other parameters and rejecting models that do not
reproduce the gravitational moments.

Results obtained for Jupiter by only fitting $R_{\rm eq}$ and $J_2$
are shown in Fig.~\ref{fig:boites-jup-reqj2}. What is particularly
striking is that there is very little overlap between the solutions
for the different EOS and that the result for the basic interior
properties, $M_{\rm core}$ and $M_{\sss Z}$, greatly depends on the
choice of EOS. Note for example that the LM-A and the LM-B EOS
predict identical Hugoniots, but yield very different
interiors. Quantitatively, this is due to the fact that the LM-B
adiabat is $\sim 10$\% less dense than the LM-A adiabat between 0.2 
and 15$\,$Mbar (Fig.~\ref{fig:adiab_Prho}).

The SESAME EOS combination of a low compressibility for $0.01<P<3\,$Mbar and
a high compressibility at larger pressures pushes the core mass 
downward. Suprisingly, even models with $M_{\rm core}=0$ do not fit
Jupiter's $R_{\rm eq}$ and $J_2$.  This probably
means that the SESAME EOS is too crude, but alternatively, that Jupiter's
interior might be more complex than assumed here. A very small
set of models with $M_{\rm core} \sim 1\,M_\oplus$ and $M_{\sss Z}
\sim 33\,M_\oplus$ were found with the SESAME-p EOS (which was
softened in the low-pressure molecular region to be consistent
with low-pressure experimental data).

Figure~\ref{fig:boites-pure-jup} shows the same results but including
the constraint on $J_4$. A noticeable consequence is that LM-B EOS no
longer satisfies the constraints. In fact, the values of $J_4$ found
with this EOS are always more than $2\,\sigma$ away from the observed
value (towards a higher absolute $|J_4|$). This EOS can apparently be
ruled out.
All other EOS show solution ensembles that barely differ from
Fig.~\ref{fig:boites-jup-reqj2}. In fact, models based on the SCVH-I EOS are
the only ones that can be further constrained by the value of $J_4$,
as the upper-right part of the box in Fig.~\ref{fig:boites-jup-reqj2}
corresponds to values of $J_4$ which lie 1 to $2\,\sigma$ away from
the observed value. The absence of any further change for the other
EOS shows that most of the constraints on Jupiter's interior stem
from the inferred radius and $J_2$. The measured $J_4$ is useful, but
it is not known with a sufficient accuracy yet to play an important
role in the determination of $M_{\rm core}$ and $M_{\sss Z}$. 

In order to estimate the robustness of the $(M_{\rm core},M_{\sss Z})$
ensemble of solutions for a given EOS, we considered models in which the density profile
along the adiabat is arbitrarily modified by $2\%$ according to:
\begin{equation}
\rho(P)=\rho_0(P) \left\{1\pm 0.02\,e^{-\left[\log_{10}(P/P_0)\right]^2}\right\},
\end{equation}
where $\rho_0$ is the unperturbed density and $P_0$ is varied from 3 to 100\,Mbar. 
This provides an estimate of the remaining uncertainty on the adiabat computed from
a given H EOS model and accounts for the approximate nature of the additive volume 
rule that we use to generate the EOS of the mixture of hydrogen, helium, rocks, and ices.
Figure~\ref{fig:boites-jup} shows that the resulting core masses and
mass of heavy elements in the envelope globally lie relatively close
to the solutions calculated without that additional
uncertainty. In particular, the modification is not sufficient to
salvage the LM-B EOS, whose resulting models still have $J_4$ values
that are outside the $2\sigma$ observational error bars. 

Quantitatively, we find that in Jupiter, $0 \le M_{\rm core}
\wig< 11\,M_\oplus$ and $1 \wig< M_{\sss Z} \wig< 39\,M_\oplus$.  The
total amount of heavy elements $M_{\rm core} + M_{\sss Z}$ is 8 to
39$\,M_\oplus$.  Compared to the abundance of heavy elements in the
Sun, this corresponds to an enrichment by a factor of 1.5 to 6, which
is consistent with the enrichment observed in Jupiter's atmosphere.
Since the full range of solutions is quite a bit larger than allowed
for any given EOS, it implies that the uncertainty associated with the
hydrogen EOS is more significant than all of the other uncertainties
taken together.  Furthermore, it is not yet possible to determine with
any confidence the mass of the core of Jupiter or whether it has a
core at all!  Similarly, the total content of heavy elements is even
less constrained.  Only the LM-A EOS predicts a core mass as high as
11$\,M_\oplus$, which is marginally consistent with the core accretion
formation model (which normally requires 10 -- 20$\,M_\oplus$).  All
the others predict cores at most half as massive.  

A comparison of Figs. 4, 5 and 8 shows that the hydrogen EOS that produce 
softer principal Hugoniots (i.e. 
reach higher maximum compression) give models with somewhat larger $M_{\rm core}$ and 
smaller $M_{\sss Z}$, Further explorations with additional hydrogen EOS 
reveal that this trend is only apparent, however.  Stiff EOS can lead to value of $M_{\sss Z}$
nearly as small as those from the softest EOS. In other words, the EOS
{\it model} fitted to the experimental data used to compute
the adiabat is just as important as the experimental data themselves
in determining the internal structure of Jupiter.  This points to the 
importance of measuring the EOS of hydrogen closer to the Jupiter isentrope.

\subsection{Jupiter: Uncertainties related to the He EOS}

In order to estimate uncertainties due to the helium EOS, we
calculated Jupiter models with the same hydrogen EOS as described
previously, but using the He-SESAME-p EOS instead of the He-SCVH EOS
(see section~II.C). The two EOS differ only slightly, but the
systematic deviation in both entropy and density lead to a
$\sim 2\%$ reduction in density along the adiabat of the mixture
for $P\wig{>}3\,$Mbar.

The consequences are shown in Fig.~\ref{fig:boites-jup-he}. With the new
He-SESAME-p, the core mass is found to be systematically larger. This is
explained by the fact that the $2\%$ reduction in the density of the mixture
needs to be
compensated by an additional $\sim 2\%\times 318\,M_\oplus \sim 6\,\rm M_\oplus$ mass
of heavy elements, to be distributed between $M_{\rm core}$ and $M_{\sss Z}$.
Another consequence is a displacement of the resulting value of $J_4$,
for any model fitting $R_{\rm eq}$ and $J_2$ (see Fig.5 of
Guillot \citep{guillot99}). The new values are such that $|J_4|$ is
higher by about $0.5-1\sigma$. This implies that solutions with the
LM-B hydrogen EOS are even further from the observations than
before. It has little consequences for other EOS, except for the
SCVH-I solutions which now always have a $J_4$ value that is at least
$1\sigma$ away from the measurements.  Although the solutions set
by $R_{\rm eq}$ and $J_2$ have moved to higher values of $M_{\rm
core}$ and $M_{\sss Z}$, the $J_4$ constraint reduces the size of the effect.

\subsection{Saturn}

The mass of Saturn is slightly below $1/3$ that of Jupiter and as a
consequence, only 67\% of its mass lies at $P>1\,$Mbar, compared to
91\% for Jupiter.  Saturn is therefore less sensitive to the larger
EOS uncertainties than Jupiter.  Saturn's response to rotation is also
qualitatively different, as can be seen from its linear response
coefficient: $\Lambda_2\approx J_2/q$ where $q=\omega^2R_{\rm
eq}^3/GM$ is the ratio of the centrifugal to the gravitational
potentials. The value of $\Lambda_2$ is $0.108$ for Saturn compared to
$0.166$ for Jupiter, implying that Saturn's interior is more
concentrated, i.e. it is expected to have a proportionally larger core than
Jupiter \citep{hubbard89}. Finally, the error bars on the
gravitational moments $J_2$ and $J_4$ of Saturn are much larger than
for Jupiter.  We anticipate that the results will be
qualitatively different.

The solutions for Saturn are summarized in Fig.~\ref{fig:boites-sat}.
We present only the case where the 2\% perturbation on the EOS (Eq. 7)
has been included as this has only a slight effect on Saturn models.  We find
that there is much more overlap between the solutions for the
different EOS than is the case for Jupiter and that we can usefully
constrain the core mass and the mass of heavy elements in the envelope
to $9 \wig< M_{\rm core} \wig< 22\,M_\oplus$ and $1 \wig< M_{\sss Z}
\wig< 8\,M_\oplus$ for a total mass of heavy elements between 13 and
28$\,M_\oplus$.  The EOS is an important source of uncertainty in
modeling Saturn's interior but it is not as dominant as in Jupiter.
We also find that the LM-B EOS and the original SESAME EOS (not shown)
give acceptable models of Saturn. The amount of heavy elements in the
envelope is rather modest but the total amount of heavy elements in
Saturn represents a 6- to 14-fold enrichment compared to the solar
value.  Saturn may contain more heavy elements than Jupiter.  
The choice of EOS has no discernible effect on $M_{\rm core}$, however,
and the uncertainty on the
helium EOS barely affects the ensemble of solutions for Saturn.
The distribution and the amount of heavy elements in Saturn is in very
good agreement with the core accretion formation model.

\section{Evolution of Jupiter}

While the $(P,\rho)$ relation along the adiabat determines the
internal structure of the planet, the $(P,T)$ relation influences its
cooling age by setting its internal heat content.  The cooling rate
itself is determined by the rate at which heat can escape at the
surface, which is controlled by the atmosphere of the planet, treated
as a surface boundary condition in the cooling calculation.  In
section 4, we showed how rather
modest variations in density along the adiabat
(Fig.~\ref{fig:adiab_Prho}) are responsible for astrophysically
significant changes in internal structure.  The variations in adiabat
temperatures between the EOS can be as large as 60\%.  The largest
variations occur at pressures above 1$\,$Mbar and involve most of the
mass of Jupiter and Saturn.

In order to test the effect of the EOS on the evolution of the giant
planets, we have computed homogeneous, adiabatic evolution calculations
(i.e. without considering the possibility of a He/H phase separation)
for Jupiter following the method described in \citet{gcgm95}. Although
the resulting models only reproduce approximatively the present
planetary structures (we only try to match the mean radius and intrinsic
luminosity), it is a helpful indication of how the hydrogen EOS also
affects the cooling of the giant planets. We do
not present calculations for Saturn since its evolution is likely
affected by the slow gravitational energy release from
helium-rich droplets sinking toward the center due to H/He demixing 
\citep{fortney03}.

For each choice of EOS, initial
conditions are determined by the structural parameters fitted to the
observational constraints.  The model is started at an arbitrary large
value of the specific entropy, resulting in an initial radius
$R\approx 1.15-1.5\,R_{\rm Jup}$. Because the planet initially cools
very rapidly, the initial value of the entropy is not important as
long as it is large enough (the corresponding uncertainty between the
different models is $\sim 50\,$Myr). The variation in internal
heat content with the choice of EOS implies that Jupiter will cool to
its present luminosity (and contract to its present radius) in more or
less time.  This cooling age is to be compared to the age of the 
Solar System of 4.56$\,$Gyr.

Naturally, models computed with EOS that give cooler adiabats (such
as LM-A and LM-B, see Fig.~\ref{fig:adiab_PT}) have a lower internal
heat content and will cool to the present luminosity in a shorter
time, everything else being equal.  The effect is quite significant,
as can be seen in Fig.~\ref{fig:evol_jup03}.  We find that the cooling
ages for each EOS are $5.4\,$Gyr (SESAME-p), $4.8\,$Gyr (LM-SOCP),
$4.7\,$Gyr (SCVH-I), $4.0\,$Gyr (LM-H4), and $3.1\,$Gyr (LM-A). The
absolute ages quoted here are only representative since they depend on
a presently inaccurate surface boundary condition and do not include
the possibility of a modest degree of phase separation in Jupiter
which would lengthen the cooling time of all models \citep{fortney03}.
The relative ages are much less sensitive to these limitations.

Nevertheless, the cooling age obtained with the LM-A EOS is so short
that it is unlikely that it could be reconciled with the age of the
solar system with a more elaborate cooling calculation.  Adiabats that
are very cool above 1$\,$Mbar are characteristic of the EOS that fit
the reshock temperature data (LM-A, LM-B, and the even more extreme
case of the original LM model of Ross shown in
Fig.~\ref{fig:J_ad_Ross}).
On the other hand, it appears that the SESAME-p EOS leads to a cooling
that is too slow to be consistent with the age of the Solar
System. For the other EOS (LM-H4, SCVH-I and LM-SOCP), it can be
advocated on one hand that energy release due to a H/He phase
separation may lengthen the cooling \citep{fortney03}, or on the other
hand an efficient erosion of the central core may shorten
it sufficiently to account for the observed discrepancy \citep{gshs03}.

\section{Astrophysical implications and concluding remarks}

We have computed a new generation of interior models for Jupiter and
Saturn, with an emphasis on improving the treatment of the EOS of
hydrogen and of elements heavier than H and He. In particular, we have
used 7 different EOS of hydrogen that were chosen to reproduce the
range of possibilities indicated by first and second shock Hugoniot
data.  This allows, for the first time, a determination of the effects
of the present uncertainty on the EOS of hydrogen on the interior
structure of jovian planets. The parameters of interest, which
characterize the total mass of heavy elements and their radial
distribution in both Jupiter and Saturn are now constrained reliably
by this study.  

It is interesting to compare the results to previous
models \citep{ggh97, guillot99} for which a third parameter was added
to the optimization: a discontinuity in the abundance of heavy
elements accross the helium-poor/helium region 
(located at the transition between molecular and metallic hydrogen
in the 3-layer model). 
Our results for Jupiter are very similar, in particular concerning the
low central core mass and highly uncertain total mass of heavy
elements. Differences arise in the case of Saturn, for which some
models \citep{guillot99} were found to fit the planet's
gravitational field with no central dense core. However, these
previous models consisted in solutions with a high abundance
discontinuity at the molecular/metallic transition (a large abundance
of heavy elements in the helium-rich region mimicking the effect of a
central dense core).
We can thus be relatively confident that even with a simple
three-layer model with two free parameters ($M_{\rm core}$ and $M_{\sss Z}$)
one can constrain the global structures of Jupiter and Saturn. On the
other hand, we should keep in mind that the real structures of these two
planets maybe more complex so that the actual interpretation of these
parameters is not precisely defined (e.g. the central core could be
either diluted or present as a well-defined structure).  

Quantitatively, we confirm that both Jupiter and Saturn are enriched
in heavy elements compared to the Sun, by factors of 1.5 -- 6 and 6 --
14, respectively.  Maximum
compression ratios of $\sim 4$ along the principal Hugoniot are
supported by three independent experiments \citep{knudson01, belov02,
boriskov03, knudson03, knudson04}, and by most ab initio EOS simulations
(Lenosky, Kress \& Collins 1997; Militzer \& Ceperley 2000; Galli et al. 2000;
Desjarlais 2003).  
This type of
Hugoniot response is represented by the LM-SOCP and the SESAME-p
EOS. Interestingly, both EOS lead to very small core masses ($M_{\rm
core} \wig< 3\,M_\oplus$) in Jupiter, and a substantial amount of
heavy elements in the envelope.  The same EOS predict a more massive
core in Saturn (10 -- 20$\,M_\oplus$).  Changing the He-SCVH helium
EOS for the He-SESAME-p EOS
has a modest effect on the structure of Jupiter and increases the total 
amount of heavy elements by a few Earth masses, split more or less
evenly between the core and the envelope.  This does not change the
qualitative picture that emerges, however.

Taken at face value, these results imply
that Jupiter formed by disk instability in the protoplanetary
disk while Saturn formed by core accretion.  It is rather unlikely
that the two planets formed by such different mechanisms, however.  We
speculate that the only way to reconcile the formation processes of
Jupiter and Saturn is that both of them formed by core accretion and
that for Jupiter, the subsequent accretion of the gaseous
H/He envelope resulted in partial or complete mixing of the core with 
the gas, increasing
$M_{\sss Z}$ at the expense of $M_{\rm core}$.  The larger accretion
rate in the proto-Jupiter may have caused larger mixing than in
the proto-Saturn \citep{gshs03}.

On the other hand, Jupiter models computed with the LM-A EOS have core
masses that are (barely) consistent with the mass required for
formation with core accretion formation ($\sim 10\,M_\oplus$).  In
this case, mixing of the core of proto-Jupiter with the accreting
envelope would not be required and both planets would form by the same
process.  The LM-A EOS was constructed to fit the NOVA $(P,V,T)$
Hugoniot as well as the gas gun reshock temperatures.  It represents
the softest and coolest EOS allowed by the experiments.  The reshock
temperatures measurements may be too low, however, and a reevaluation
of these measurements may well rule out the LM-A EOS.

This study has been conducted with the point of view of learning about
the interiors of Jupiter and Saturn from our current knowledge of the
EOS of hydrogen and the associated uncertainties.  The opposite
approach can also be considered: Can astrophysical knowledge
contribute to the debate on the high-pressure EOS of hydrogen?
Because much astrophysical knowledge is not amenable to direct
observation or experimentation, our knowledge of processes that are
hidden from view or that are no longer taking place is very sketchy.
It is usually difficult to draw strong conclusions about the
underlying microphysics when only the global properties of a complex,
natural object are known.  The ability of a given EOS to give
acceptable models depends somewhat on the assumptions for the model
structure.  In general, more elaborate models (with more parameters)
can accommodate a wider range of EOS. Nevertheless, we venture to
comment on interesting patterns that emerge in the more extreme cases
that we have encountered in this study.

\begin{figure}
\plotone{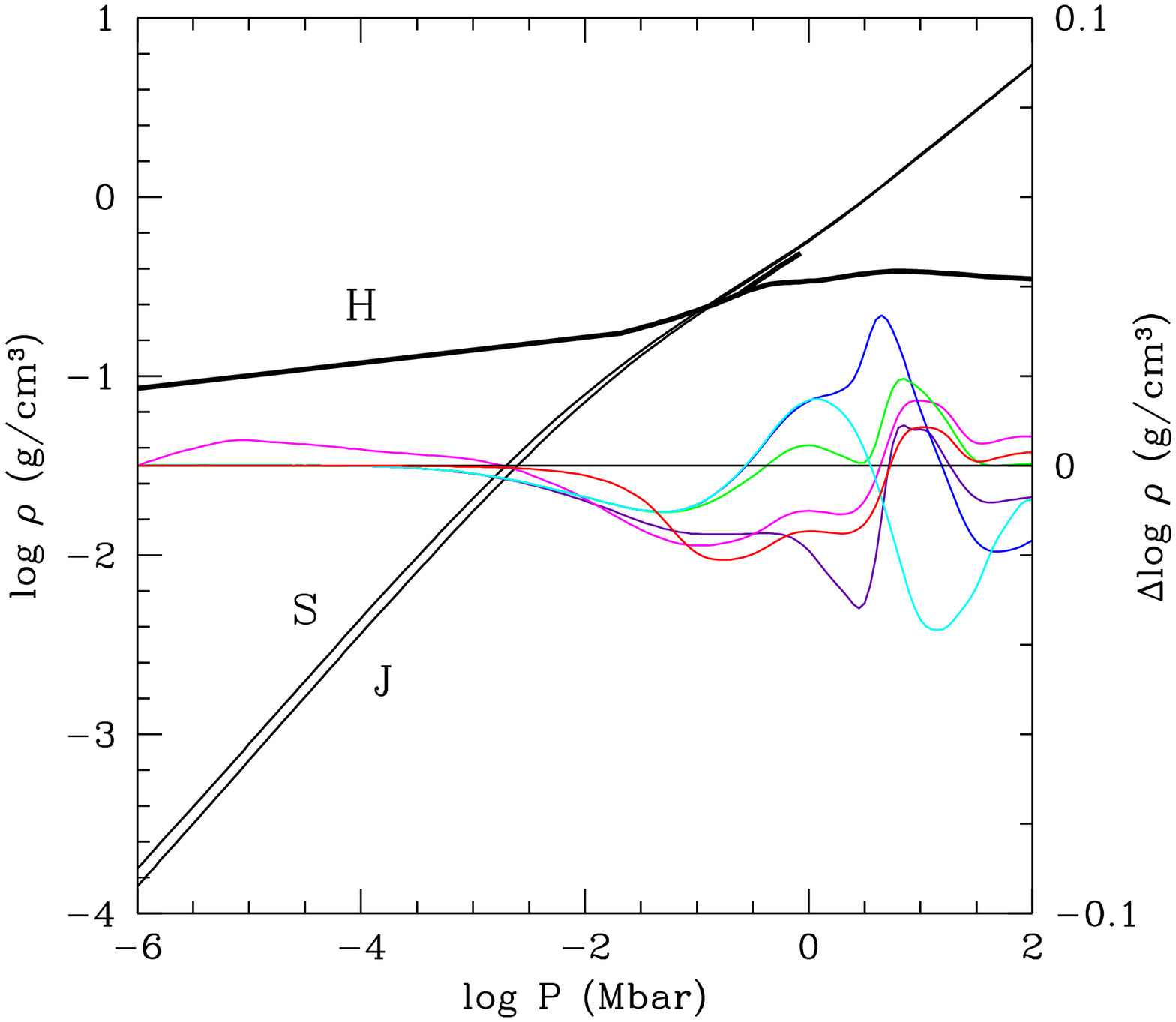}
  \caption{Adiabats for hydrogen in $P$--$\rho$.  The curves labeled
`S' and `J' show the SCVH-I EOS adiabats of Saturn and Jupiter,
determined by $T=136\,$K and 170.4$\,$K at $P=1\,$bar, respectively.
The first and second-shock Hugoniots (SESAME EOS) are shown by the
heavy solid line labeled `H'.  The other curves and the scale on the
right show {\it differences} in density between Jupiter adiabats
computed with various EOS (see section
2.A), relative to the SCVH-I:
SESAME (short dashes, magenta), SESAME-p (short dash - long dash, red), LM-A
(long dashes, blue), LM-B (long dash - dot, cyan), LM-SOCP (dotted, purple), and LM-H4
(short dash - dot, green).  The central pressure of Jupiter is about
70$\,$Mbar and that of Saturn is about 40$\,$Mbar.
[{\it See the electronic edition of the Journal for a color version of this figure.}]}
\label{fig:adiab_Prho}
\end{figure}

\begin{figure}
\plotone{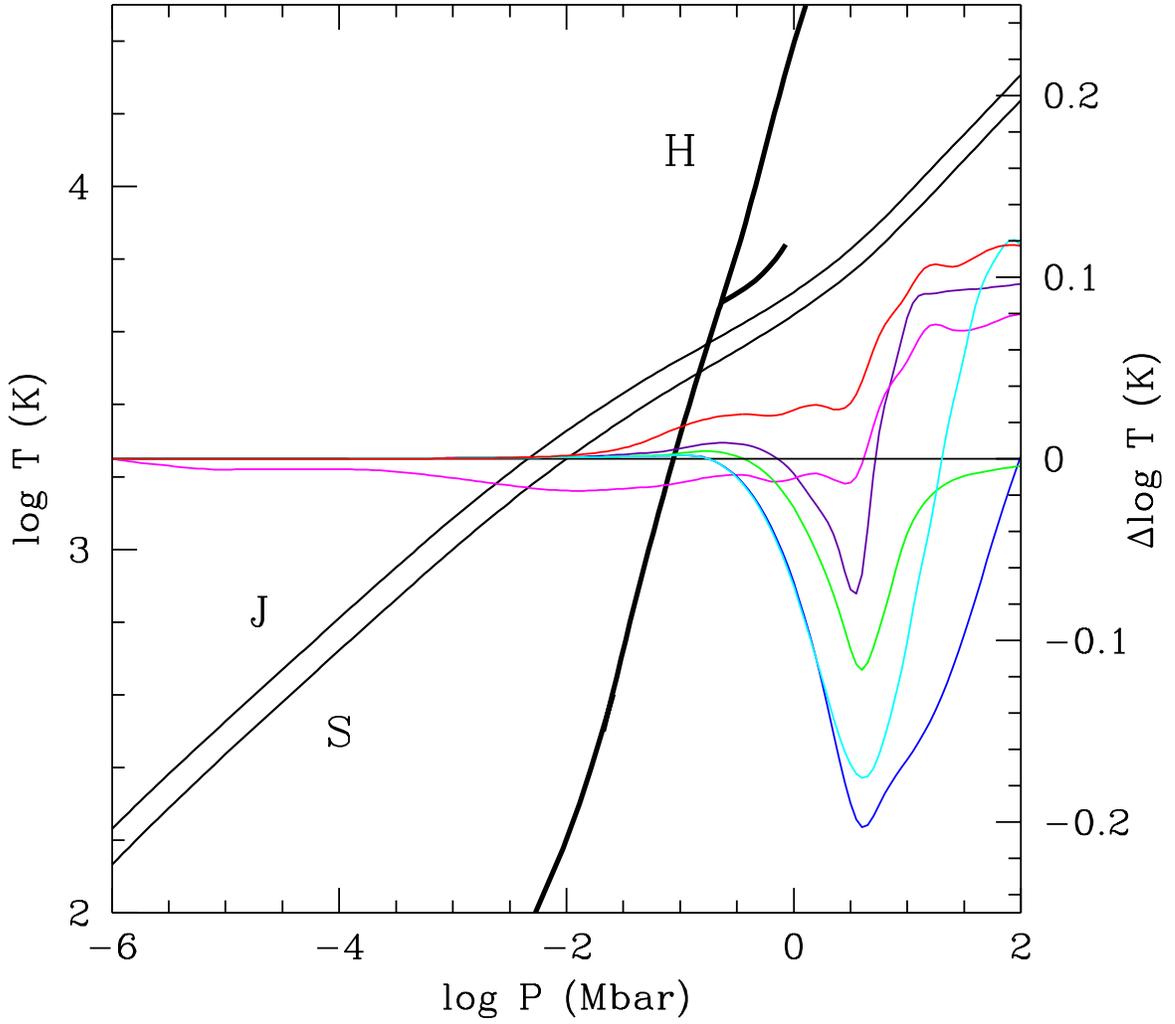}
  \caption{Adiabats for hydrogen in $P$--$T$.  See the caption of
  Fig.~\ref{fig:adiab_Prho} for details.
[{\it See the electronic edition of the Journal for a color version of this figure.}]}
\label{fig:adiab_PT}
\end{figure}

\begin{figure}
\plotone{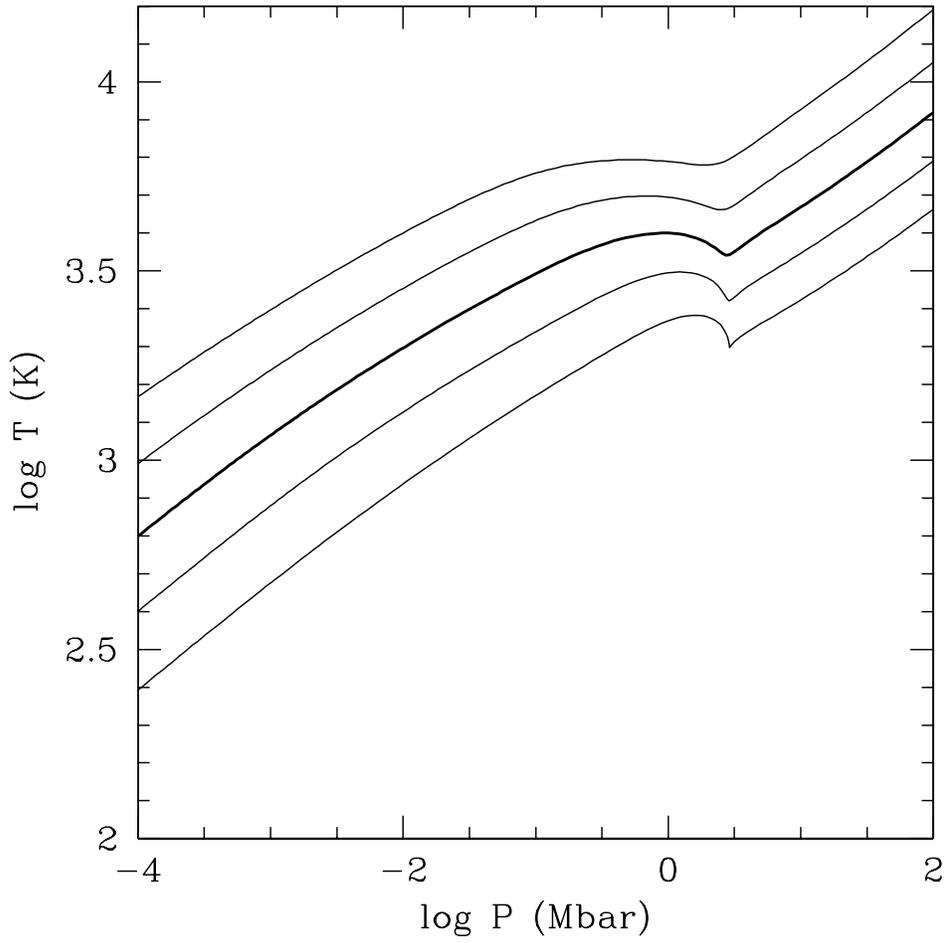}
  \caption{Adiabats for hydrogen computed with the linear mixing model
           of \citet{ross98} for different specific entropies.
           The entropy decreases from top to bottom.  Jupiter's
           adiabat is shown by the heavy solid line.}
\label{fig:J_ad_Ross}
\end{figure}

\begin{figure*}[!]
\plotone{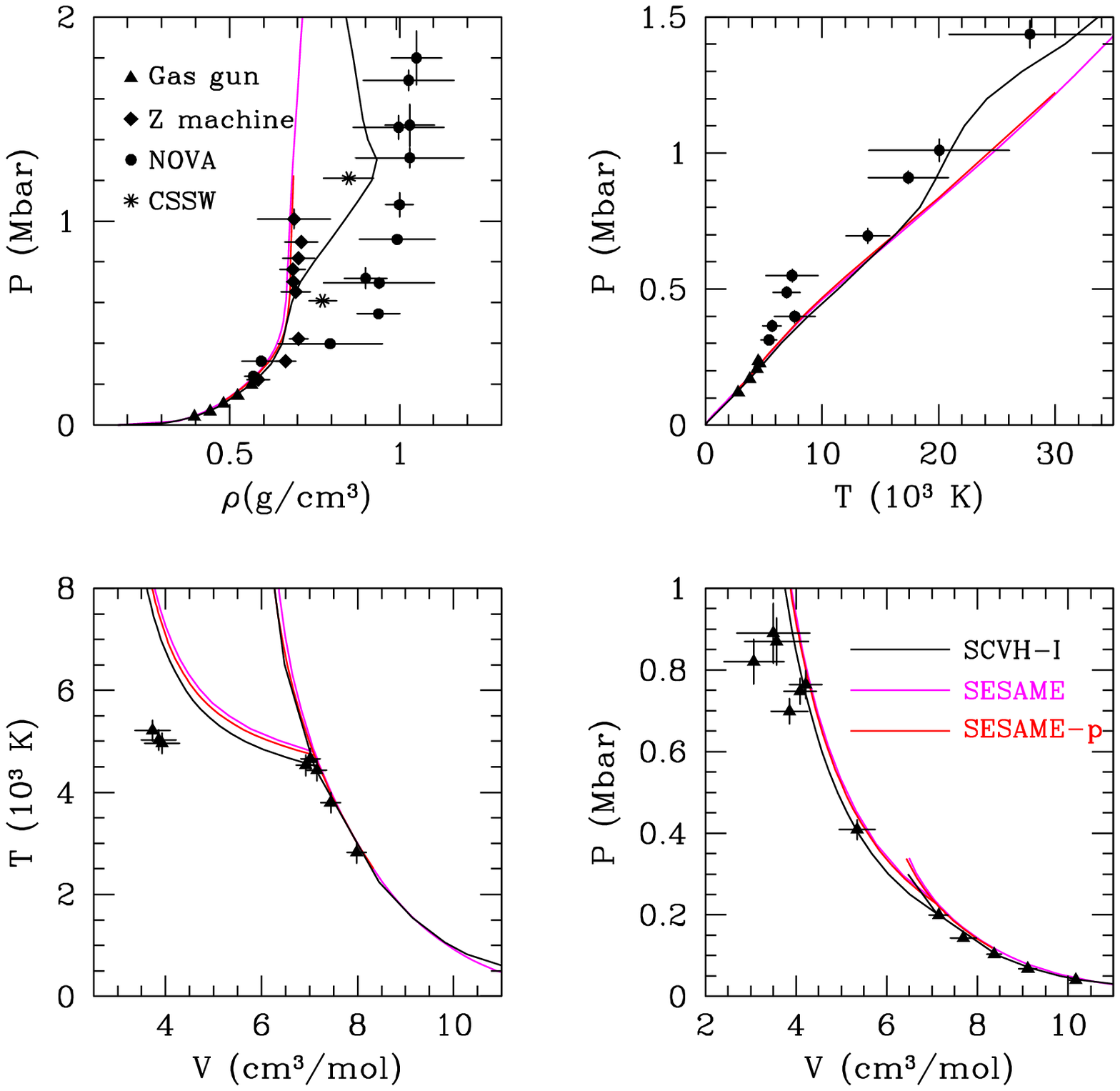}
  \caption{First and reshock Hugoniots of deuterium compared with
           experimental data. The Hugoniots are computed from the
           following EOS: SCVH-I, SESAME, and SESAME-p.  Both
           versions of the SESAME EOS give very nearly identical
           Hugoniots.  The experimental data includes gas gun 
           data \citep{vanthiel74, nellis83, holmes95}, the NOVA 
           data \citep{dasilva97, collins98, collins01}, Z-machine
           measurements \citep{knudson04} and one point from a
           convergent spherical shock wave compression 
           experiment \citep{belov02, boriskov03}.
[{\it See the electronic edition of the Journal for a color version of this figure.}]}
\label{fig:hugo_panels_A}
\end{figure*}

\begin{figure*}[!]
\plotone{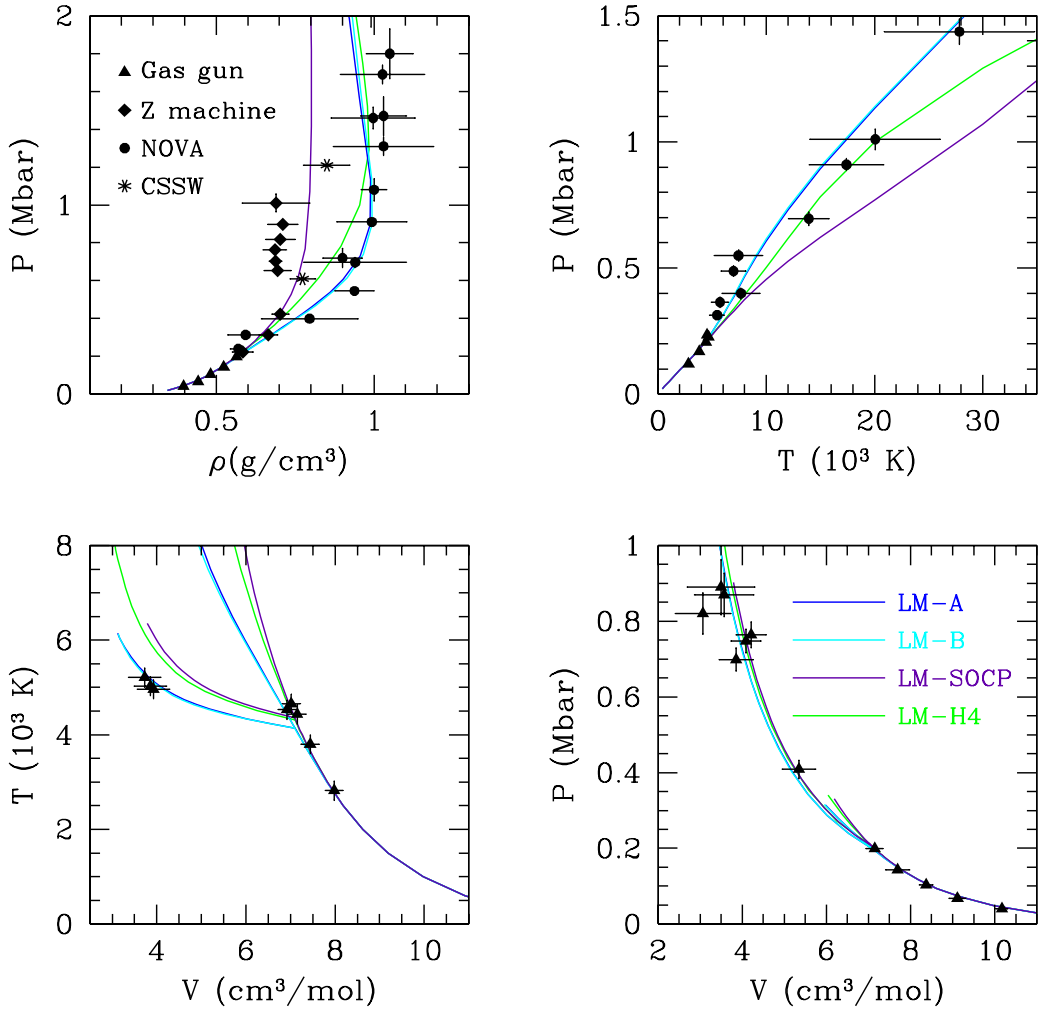}
  \caption{Same as Fig.~\ref{fig:hugo_panels_A} but for the following
           EOS: LM-A, LM-B, LM-SOCP, and LM-H4. The LM-A and LM-B
           Hugoniots are very nearly identical.
           [{\it See the electronic edition of the Journal for a color version of this figure.}]}

\label{fig:hugo_panels_B}
\end{figure*}

\begin{figure}[!]
\plotone{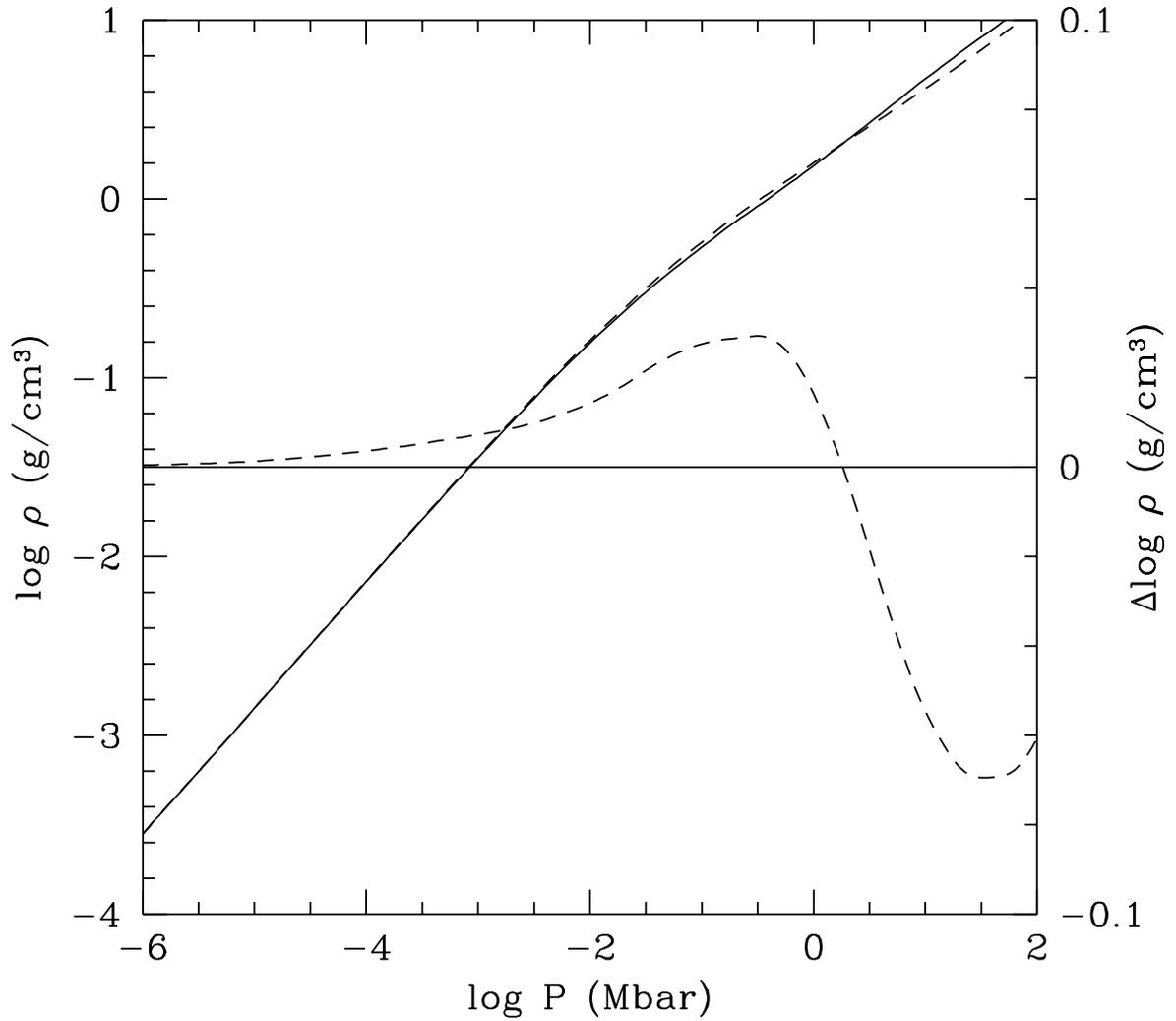}
  \caption{Density of helium along the $(P,T)$ adiabat of Jupiter for
           hydrogen (diagonal curves).  Two EOS are shown: He-SCVH (solid) and 
           He-SESAME-p (dashed). 
           The difference between the two adiabats is shown on the scale at right.}
\label{fig:He_adiab}
\end{figure}

\begin{figure}[t]
\resizebox{1.0\columnwidth}{!}{\includegraphics[angle=-90]{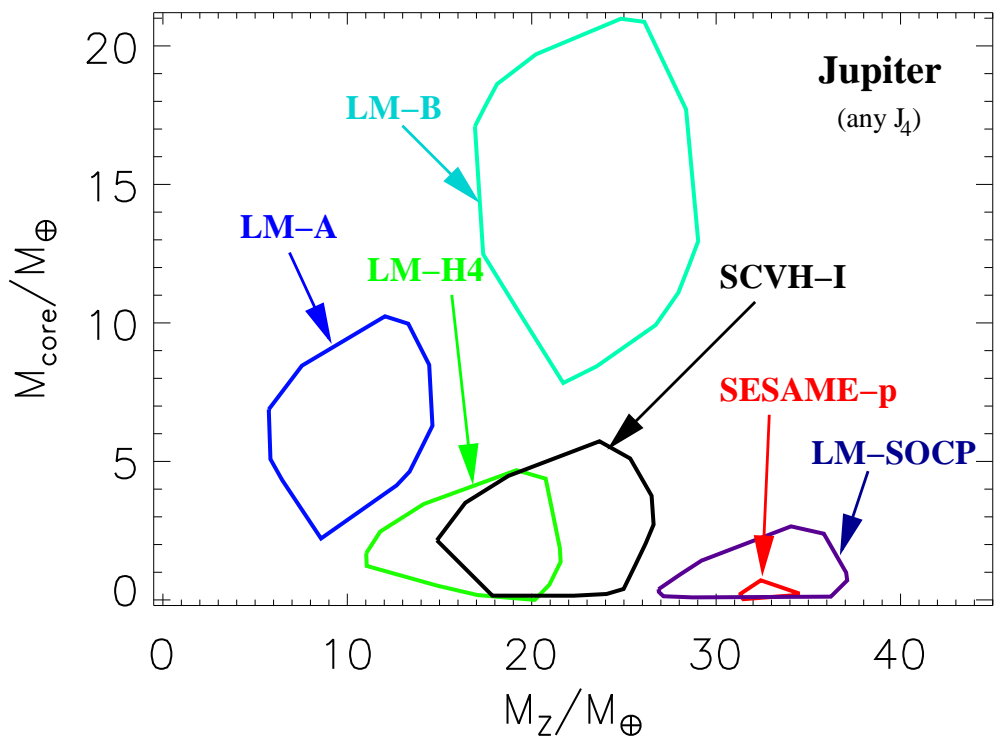}}
  \caption{Jupiter's core mass $M_{\rm core}$ and mass of heavy
elements mixed in the H/He envelope $M_{\sss Z}$ in Earth masses
($M_\oplus$). The total mass of Jupiter is $317.83\,M_\oplus$. Each box
represents the range of models that match Jupiter's equatorial radius
and gravitational moment $J_2$ for a given choice of hydrogen EOS,
including all the parameter variations described in the text. The
models do not necessarily have the correct value of $J_4$.
[{\it See the electronic edition of the Journal for a color version of this figure.}]}
\label{fig:boites-jup-reqj2}
\end{figure}

\begin{figure}[!]
\resizebox{1.0\columnwidth}{!}{\includegraphics[angle=-90]{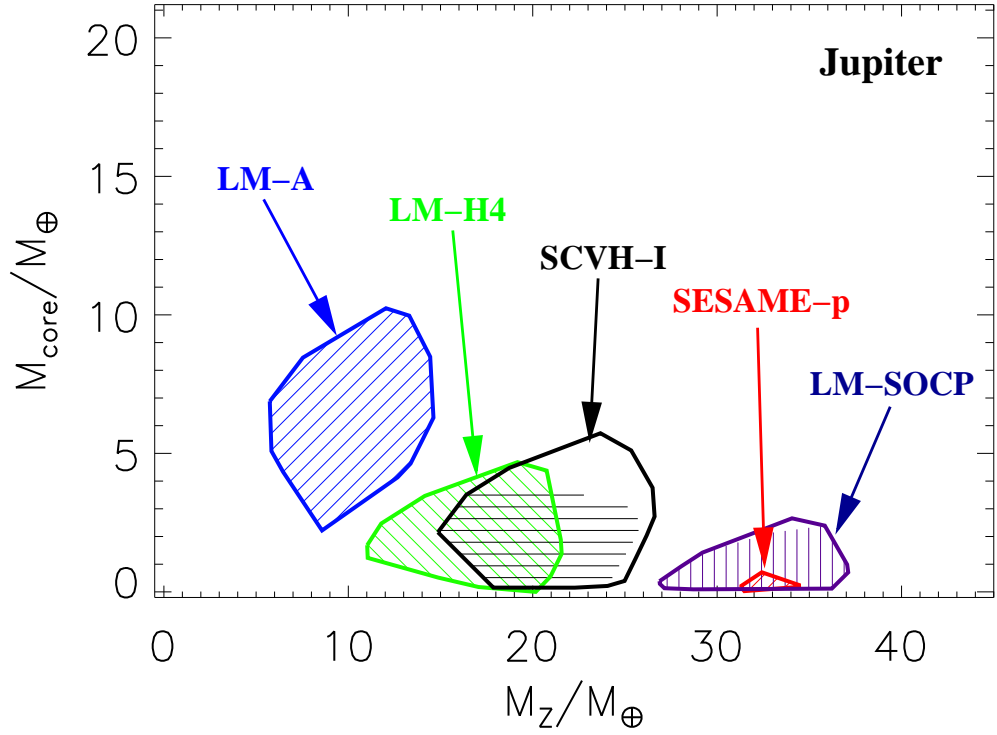}}
  \caption{Same as Fig.~\ref{fig:boites-jup-reqj2} but each box
  represents the range of models that match Jupiter's $R_{\rm eq}$,
  $J_2$ and $J_4$ within $2\,\sigma$ of the observed value. The hashed
  regions corresponds to models that match  $J_4$ within $1\,\sigma$ of the
  observations.  [{\it See the electronic edition of the Journal for a color 
  version of this figure.}]}
\label{fig:boites-pure-jup}
\end{figure}

We could not obtain satisfactory models of Jupiter with the original
SESAME EOS because it is relatively stiff between 0.1 and 3$\,$Mbar,
and relatively soft at higher pressures along Jupiter's adiabat.  Both
of these effects combine to decrease the core mass \citep{guillot99}
but even with $M_{\rm core}=0$ models that fit the gravitational
moments could not be obtained with this EOS.  It is only after it was
patched at low pressures and that a 2\% uncertainty in the EOS was
introduced that acceptable models could be found (SESAME-p EOS).
Similarly, we could not obtain satisfactory models of Jupiter with the
LM-B EOS because it is too stiff along the adiabat above 5\,Mbar.
Taken together, these constrain the $(P,\rho)$ relation along the
adiabat a little more than the experimental results alone (see
Fig.~\ref{fig:adiab_Prho}).  We find that most EOS predict relatively
small core masses for Jupiter.
While it may be possible to accommodate such a situation
astrophysically, it would require a substantial revision of the
generally accepted formation process of the planet.  Cooling
calculations provide additional information.  We find that EOS that
fit the gas gun reshock temperatures (e.g. LM-A) give cooling times that are
uncomfortably short for Jupiter, suggesting that those temperatures
are systematically too low.  Models of Saturn do not provide any
useful information on the EOS of hydrogen.

\begin{figure}[!]
\resizebox{1.0\columnwidth}{!}{\includegraphics[angle=-90]{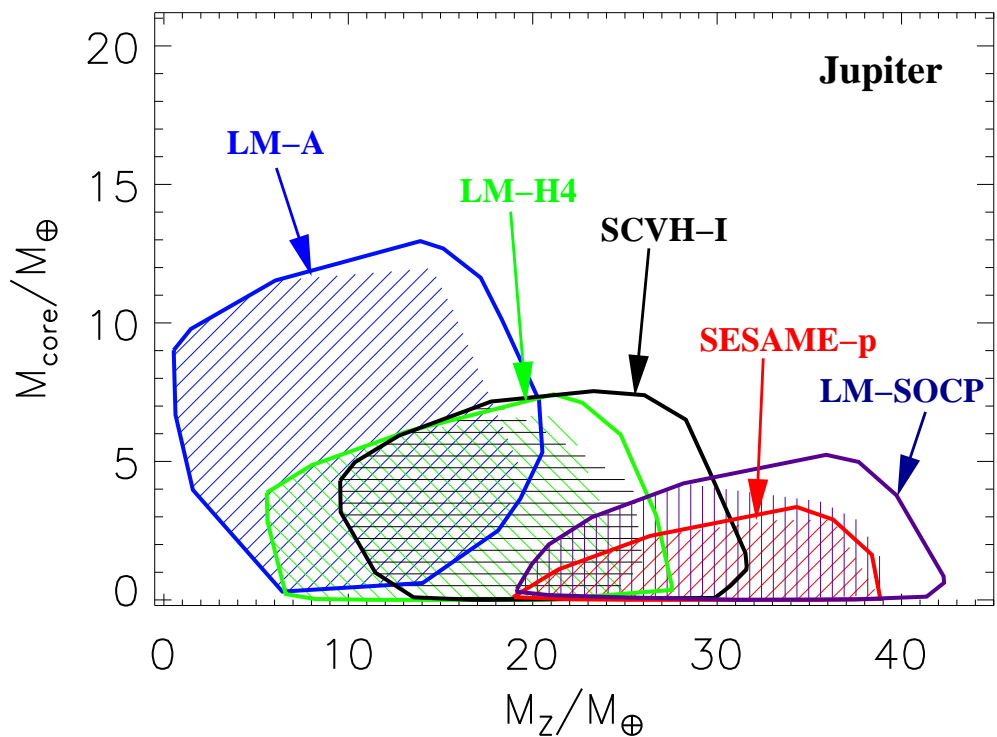}}
  \caption{Same as Fig.~\ref{fig:boites-pure-jup} but including an
  additional 2\% uncertainty on the density profile of each EOS (see Eq. 7).
  The hashed regions represent solutions for which $J_4$ is
  within $1\sigma$ of the measured value.  [{\it See the electronic edition 
  of the Journal for a color version of this figure.}]}
\label{fig:boites-jup}
\end{figure}

\begin{figure}[!]
\resizebox{1.0\columnwidth}{!}{\includegraphics[angle=-90]{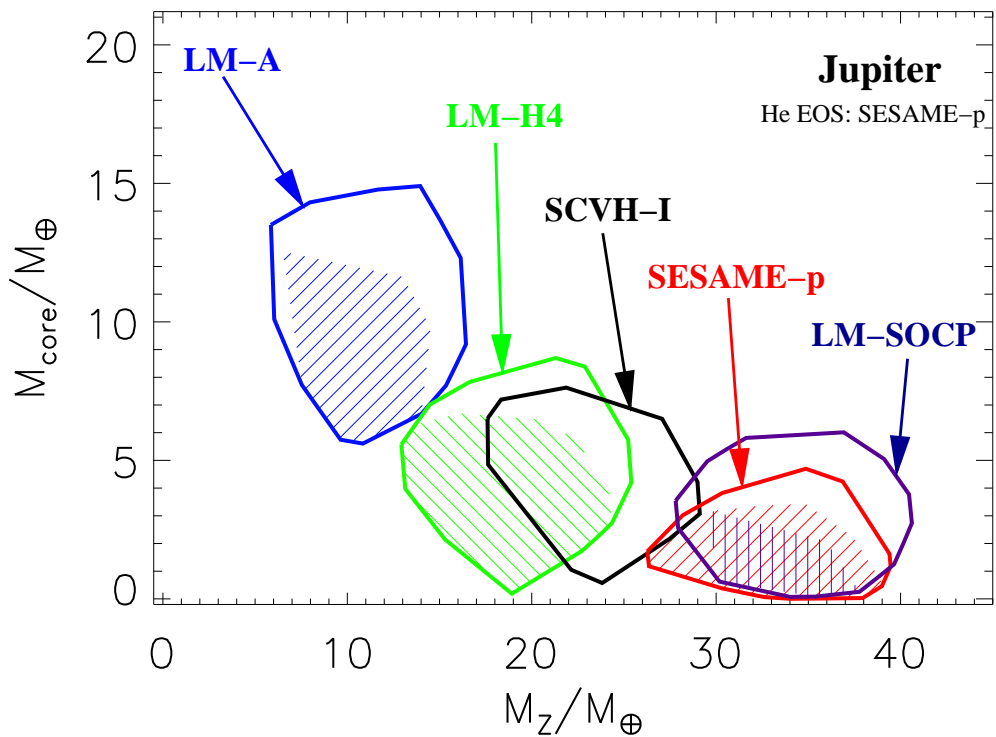}}
  \caption{Same as Fig.~\ref{fig:boites-pure-jup} but using the
  He-SESAME-p EOS for helium instead of the He-SCVH EOS (see text). The
  hashed regions represent solutions for which $J_4$ is within
  $1\sigma$ of the measured value.  [{\it See the electronic edition 
  of the Journal for a color version of this figure.}]}
\label{fig:boites-jup-he}
\end{figure}

\begin{figure}[!]
\resizebox{1.0\columnwidth}{!}{\includegraphics[angle=-90]{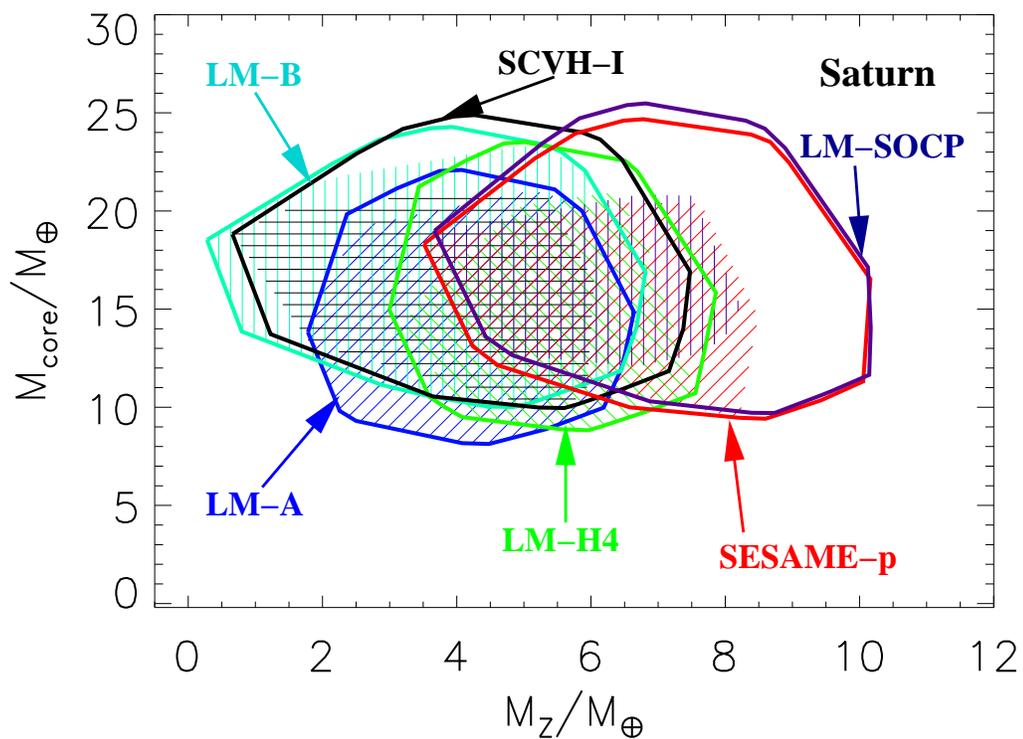}}
  \caption{Same as Fig.~\ref{fig:boites-jup} for Saturn.  Note the
           difference of scale for $M_{\rm core}$ and $M_{\sss Z}$.  A
           2\% uncertainty on each EOS is included (see Eq. 7). The
           hashed regions represent solutions for which $J_4$ is
           within $1\sigma$ of the measured value. The total mass of
           Saturn is $95.147\,M_\oplus$.
           [{\it See the electronic edition of the Journal for a color version of this figure.}]}
\label{fig:boites-sat}
\end{figure}

\begin{figure}[t]
\plotone{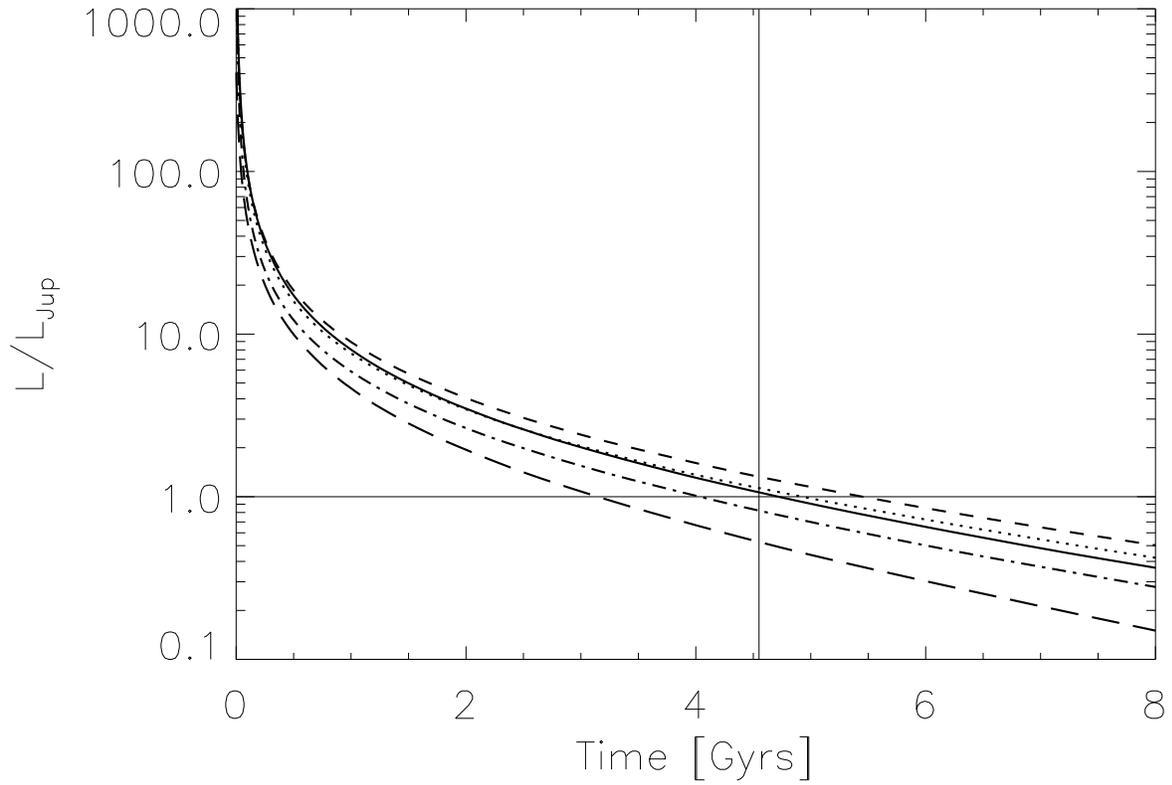}
\caption{Evolution calculations for homogeneous models of Jupiter
         using various EOS. Jupiter's present luminosity ($L_{\rm
         Jup}$) is indicated by the horizontal line. The EOS used are
         (from left to right) LM-A (long-dashes), LM-H4 (dot-dashed),
         SCVH-I (plain), LM-SOCP (dotted) and SESAME-p (dashed). The
         vertical line marks the age of the Solar System, $4.56\,$Gyr.}
\label{fig:evol_jup03}
\end{figure}

Further progress in determining the interior structures of Jupiter and
Saturn can be accomplished with a two-pronged approach.  Proposed
space missions to Jupiter would lead to a tenfold reduction of the
uncertainty on its gravitational moments, greatly reducing the range
of acceptable models ({\it i.e.} smaller boxes in
Fig.~\ref{fig:boites-jup}).  Such improved measurements would most
likely lead to rather distinct solutions for different choices of
hydrogen EOS.  The Cassini spacecraft will enter Saturn orbit in July
2004 for a 4-year mission and could reduce the error bars on the
gravitational moments significantly (in the context
of an extended mission with accurate Saturn gravity
measurements).  This would greatly reduce the 
range of acceptable models (Fig.~\ref{fig:boites-sat}).

Improved astrophysical data will not be sufficient, however.  It is as
important to reduce the uncertainties surrounding the EOS of hydrogen
in the 1 to 30\,Mbar range.  This can be achieved indirectly by reducing the
experimental uncertainty along the principal Hugoniot, or by directly
reproducing the conditions found inside giant planets with
reshock and reverberation experiments \citep{knudson03, boehly04}.
From the perspective of planetary science, two additional EOS problems 
deserve much attention.
The EOS of helium and especially that of hydrogen/helium 
mixtures, with the possiblity of a demixing phase transition,
need to be investigated both experimentally and theoretically.
The phase diagram of H/He mixtures is well-constrained
by the astrophysics of Jupiter and Saturn, making this an
especially interesting problem to study with the most advanced
ab initio methods available.

This work was supported in part by NASA Planetary Geology \&
Geophysics grant NAG5-8906, by the United States Department of Energy
under contract W-7405-ENG-36, and by the {\it Programme National de
Plan\'etologie} (France). We thank Bill Hubbard for sharing results
and codes on rotating polytropic models, and Julie Castillo for
thourough discussions on expected gravity measurements with Cassini.

\newpage

\end{document}